\documentclass{article}
\PassOptionsToPackage{numbers, compress,sort}{natbib}

\usepackage[preprint]{neurips_2025}

\usepackage[utf8]{inputenc} 
\usepackage[T1]{fontenc}    
\usepackage{hyperref}       
\usepackage{url}            
\usepackage{booktabs}       
\usepackage{amsfonts}       
\usepackage{nicefrac}       
\usepackage{microtype}      
\usepackage[table]{xcolor}
\usepackage{adjustbox}
\usepackage{enumitem}
\usepackage{wasysym}
\usepackage{amsmath}        
\usepackage{hyperref}
\definecolor{cornflowerblue}{rgb}{0.39, 0.58, 0.93}
\definecolor{lightblue}{RGB}{173,216,230}
\definecolor{lightgreen}{RGB}{144,238,144}
\definecolor{lightred}{RGB}{255,182,193}
\hypersetup{
    colorlinks=true,
    linkcolor=cornflowerblue,
    filecolor=magenta,      
    urlcolor=teal,
    citecolor=cornflowerblue,
    pdftitle={Emerging Risks from Embodied AI Require Urgent Policy Action},
    pdfpagemode=FullScreen,
    }
\usepackage{fancyhdr}

\fancypagestyle{plain}{
    \fancyhf{} 
}

\newcommand{\wcircle}{{$\RIGHTcircle$}}
\newcommand{\bcircle}{{$\CIRCLE$}}
\newcommand{\ecircle}{{$\Circle$}}

\title{Embodied AI: Emerging Risks \\and Opportunities for Policy Action}

\author{
  Jared Perlo \\
  Centre for the Governance of AI\footnote{Winter Fellow}\\
  French Center for AI Safety (CeSIA) 
  \And 
  Alexander Robey \\
  Carnegie Mellon University \\
  \And
  Fazl Barez \\
  University of Oxford \\ WhiteBox \\
  \And
  Luciano Floridi \\
  Yale University \\ 
  University of Bologna
  \And
  Jakob Mökander \\
  Tony Blair Institute for Global Change \\
  Yale Digital Ethics Center
}

\begin{document}

\maketitle

\begin{abstract}
The field of embodied AI (EAI) is rapidly advancing. Unlike virtual AI, EAI systems can exist in, learn from, reason about, and act in the physical world. With recent advances in AI models and hardware, EAI systems are becoming increasingly capable across wider operational domains. While EAI systems can offer many benefits, they also pose significant risks, including physical harm from malicious use, mass surveillance, as well as economic and societal disruption. These risks require urgent attention from policymakers, as existing policies governing industrial robots and autonomous vehicles are insufficient to address the full range of concerns EAI systems present. To help address this issue, this paper makes three contributions. First, we provide a taxonomy of the physical, informational, economic, and social risks EAI systems pose. Second, we analyze policies in the US, EU, and UK to assess how existing frameworks address these risks and to identify critical gaps. We conclude by offering policy recommendations for the safe and beneficial deployment of EAI systems, such as mandatory testing and certification schemes, clarified liability frameworks, and strategies to manage EAI’s potentially transformative economic and societal impacts.
\end{abstract}

\section{Introduction}

Embodied AI (EAI) refers to artificial intelligence (AI) systems and agents that are grounded in the physical world and learn through perception and action~\cite{paolo_call_2024, liu_aligning_2024}. EAI systems can operate across diverse environments. For example, existing EAI applications can deliver packages~\cite{li_drone-aided_2023}, patrol public spaces as security guards~\cite{qiongfang_chinese_2024}, or care for humans in intimate settings such as elder-care homes~\cite{takenaka_ai_2025, kim_ai_2025}. EAI capabilities and domains are likely to expand significantly in the coming years~\cite{garlick2024airobots, suleyman_coming_2023}. 

EAI presents both opportunities and risks for humans. EAI systems already assist people with mobility impairments in navigating the world (e.g., autonomous cars), while future systems could fill crucial agricultural or manufacturing jobs as working-age populations decline. By augmenting and complementing human labor, EAI could foster significant economic development and prosperity~\cite{atkinson_robots_nodate}. On the other hand, EAI systems can more easily cause immediate physical damage than completely virtual AI systems and may cause significant social harm as humans and EAI systems form closer connections (particularly with applications designed for companionship)~\cite{prescott_are_2021, oravec_future_2022}. See Figure 1 for a schematic comparison of classical robots, agentic AI, and EAI. 

Recent breakthroughs in AI capabilities—particularly those related to Large Language Models (LLMs) and Large Multimodal Models (LMMs)—have catalyzed unprecedented progress in EAI systems' ability to navigate and act in the physical world~\cite{kim_openvla_2024, gemini_robotics_team_gemini_2025}. At the same time, the rise of Vision-Language-Action Models (VLAs)—which cast control as next-token prediction over interleaved visual and linguistic tokens—opens the possibility for a ``ChatGPT moment'' for robotics, with sharp jumps in capability, deployment, and public awareness. Recent debuts of models like Gemini Robotics-ER, Alibaba’s Qwen2.5-VL, and NVIDIA's Isaac GR00T N1 marked significant EAI algorithmic progress, even though these models are only slowly being paired with hardware advanced enough to translate virtual capabilities into real-world actions~\cite{gemini_robotics_team_gemini_2025, bai_qwen25-vl_2025, noauthor_nvidia_2025}. In the past few months, for example, EAI systems have completed half-marathons and shown the ability to unpack groceries with little prior context~\cite{diviggiano_china_2025, noauthor_scaling_2025}, and open-source resources from industry actors like Physical Intelligence and Unitree could spur continued technical progress~\cite{noauthor_open_2025, unitree}. 

Data acquisition—traditionally a bottleneck for EAI development due to the complexity and quantity of physical-world information needed to train models~\cite{goldberg_good_2025}—is partially being addressed through open-source datasets and cross-modality approaches~\cite{collaboration_open_2025}. Simultaneously, innovations in tactile sensing, data-chunking radar, LiDAR, actuators, and power systems are expanding the potential form factors and capabilities of EAI systems~\cite{read_robot_2024, xi_recent_2024, black_real-time_2025}. Progress in physical abilities, data collection, and deployment may lower barriers to creating high-quality models about how the external world operates~\cite{amato_data_2025}. These world models involve complex perception, planning, reasoning, and memory~\cite{fung_embodied_2025}, and increasing EAI funding and research could lead to more accurate world models and positive EAI-development feedback loops. EAI research and innovation is also quickly emerging as a new frontier in geopolitical conflict, as concerns about supply chains and national industrial policy become more salient~\cite{patel_america_2025, noauthor_chinas_2025}.  

\begin{figure}
    \centering
    \includegraphics[width=1\linewidth]{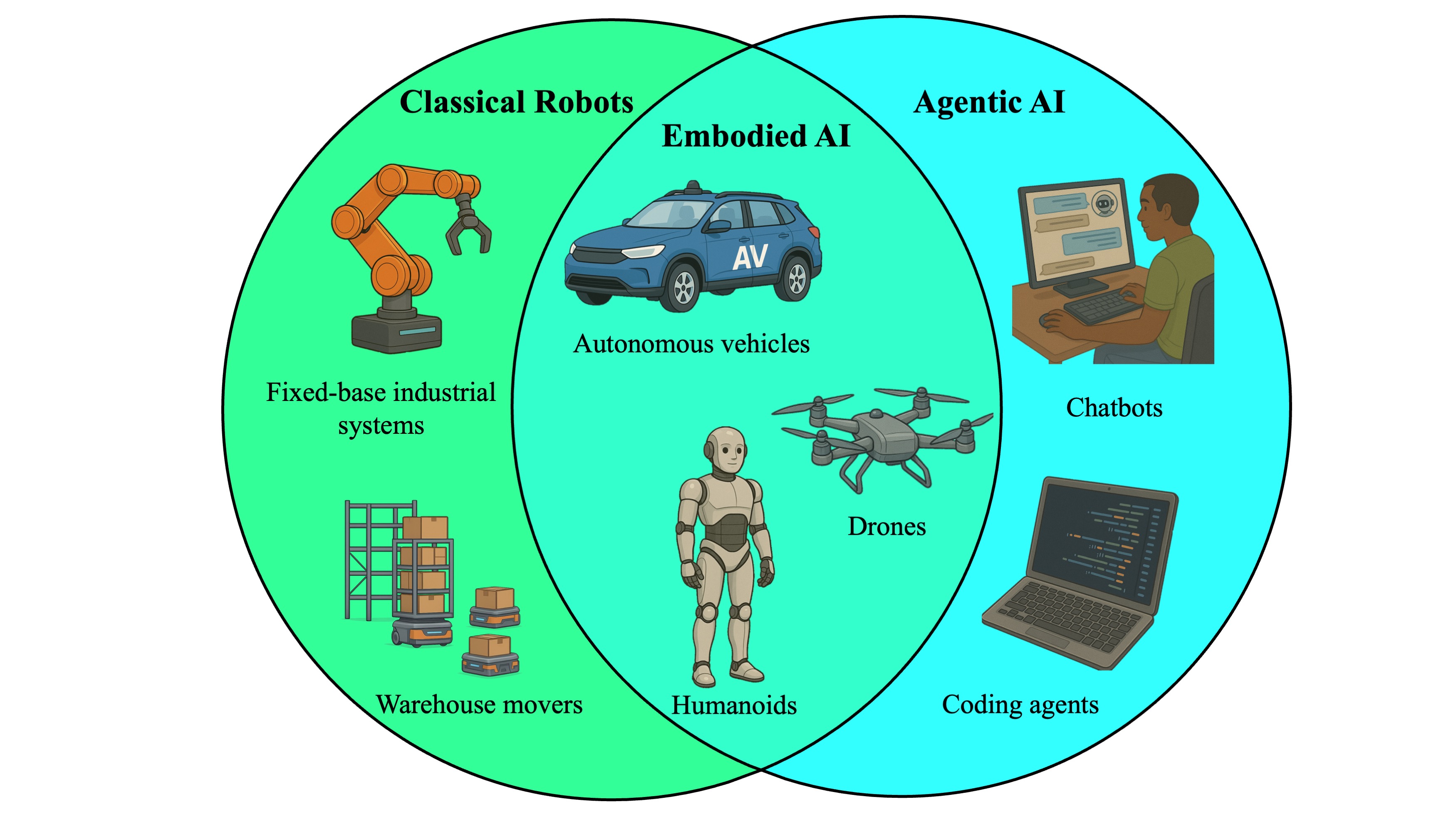}
    \caption{\textbf{A comparison of classical robots, agentic AI, and EAI} EAI represents the intersection of agentic AI and classical robots. Many existing robots, such as industrial machines (including articulated arms, gantry systems, and other types) lack the autonomy and reasoning capabilities that agentic AI possesses. Conversely, many agentic AI systems such as chatbots or virtual assistants currently lack physical embodiment.}
    \label{fig:enter-label}
\end{figure}

EAI's rapid growth in capabilities and deployment will increase the severity of potential harms and the urgency to address risks. This growth will necessitate significant updates to social, legal, and economic systems~\cite{suleyman_coming_2023, floridi_robots_2017}. Although EAI shares many characteristics with virtual agentic AI~\cite{floridi_morality_2004}, such as the varying degrees of autonomy and capability highlighted by Kasirzadeh et al. in ``Characterizing AI Agents for Alignment and Governance,''~\cite{kasirzadeh_characterizing_2025}, the physical embodiment of EAI systems introduces distinct considerations and risks that warrant special attention~\cite{xing_towards_2025}. EAI systems can hit, cut, bump, maim, attack, and more, whether intentionally or unintentionally. The complexity of the physical world presents significant adaptation challenges for digital models trained in virtual simulations~\cite{semeraro_humanrobot_2023, mazumder_towards_2023}. The coming wave of technological breakthroughs may also usher in a new era of scalability in which EAI will rapidly advance through an ever-improving software stack constrained by fewer human bottlenecks, thereby increasing the potential speed of change and compressing the timeline for action~\cite{wu_autonomy_2025}. 

Before rushing to policy action, it is essential to note that EAI is not a new concept but rather an evolution of traditional robotics. The EAI field builds upon decades of science fiction imagination, human-robot interaction research, and forecasting about advanced robotics~\cite{asimov_i_1977, barfield_cambridge_2024, akalin_taxonomy_2023}. In fact, the term ``embodied AI'' itself is partly a marketing technique used to differentiate recent innovations from traditional robotics.     

Safety concerns about EAI are likewise not novel, as researchers have studied safety in robotics for decades~\cite{zhou_essentials_1998, astrom_adaptive_1995}. Tools to formally verify robot behavior have included model predictive control~\cite{mayne_constrained_2000}, control barrier functions~\cite{ames_control_2017}, and temporal logic~\cite{kloetzer_fully_2008}, among other key innovations. Many seminal papers focusing on safe AI design prominently feature imaginary robots as examples of safe human-AI collaboration~\cite{amodei_concrete_2016, NIPS2016_c3395dd4}. More recent work has focused on creating safety guardrails for robots from real-world data, such as in Sermanet et al.'s ``Generating Robot Constitutions and Benchmarks for Semantic Safety.''~\cite{sermanet_generating_2025} However, beyond a recent UN resolution initiating discussions on lethal autonomous weapons~\cite{noauthor_general_2024}, there remains an alarming policy vacuum regarding EAI safety at national and international levels.

Understanding and minimizing risks from EAI will become even more critical in a world with AI capabilities equivalent to or surpassing artificial general intelligence (AGI), however defined~\cite{vermeer_averting_2025}. For example, increased AI-generated cyberattack capabilities could lead to perpetual attack-defense cycles, where EAI systems become targets for exploitation~\cite{shah_approach_2025}. The precise impact of AGI-level capabilities on EAI development remains uncertain, potentially accelerating deployment while simultaneously enabling more robust safety measures. AGI uncertainties aside, EAI risks are critically understudied and poorly understood, and current regulatory frameworks are generally insufficient to guide safe EAI development. 

This paper clarifies the risks and governance challenges posed by EAI and suggests a pragmatic sociotechnical approach to help governments and researchers support the development of safe EAI~\cite{watson_competing_2024}. This paper makes three unique contributions to address this urgent issue:

\begin{enumerate}[noitemsep]
	\item We develop a comprehensive taxonomy of risks from EAI, spanning physical, informational, economic, and social dimensions. This taxonomy of existing, emerging, and projected risks covers concerns ranging from malicious physical harm from jailbreaking LLMs and privacy violations in homes to widespread labor displacement. To create this taxonomy, we draw on the extensive literature related to robot safety, human-robot interaction, and recent predictions about AI's trajectory.
	\item We analyze existing policy frameworks related to EAI to assess their adequacy and highlight critical coverage gaps. Although specific pieces of legislation governing autonomous vehicles or advanced robotics trend in the right direction, significant and concerning gaps remain in existing frameworks. For example, current regulations concerning robots are ill-suited to govern systems that have high levels of autonomy and continuous learning; these characteristics challenge existing safety testing and assurance paradigms.
	\item With these risks and gaps in mind, we propose and discuss several targeted policy interventions to improve EAI safety. We suggest increasing targeted safety research, establishing robust certification requirements for EAI, promoting industry-led standards (which can offer clarity until slower-moving legislation and international agreements are passed), clarifying liability regimes, and creating substantive and actionable policy blueprints to respond to transformative economic and social effects of EAI.
\end{enumerate}

This paper has several key limitations. Given EAI's expansive and cross-cutting nature, we made practical choices to limit the scope of this paper. We primarily address civilian applications of EAI, although military and law enforcement applications also deserve special consideration. We focus on frameworks from the US, UK, and EU, though emerging regulatory efforts in other regions—especially in China—deserve increased attention and analysis. We also recognize that ensuring safe embodied AI will require a multi-layered approach, with mechanisms to enhance safety at the model, application, and organizational layers~\cite{mokander_auditing_2024}. While it remains crucial to ensure the safety of underlying models through growing AI safety research, we focus on strengthening safety measures for EAI-specific applications and organizational deployments. We likewise acknowledge that many risks from EAI are extensions of risks from powerful virtual or non-embodied AI systems. However, EAI also presents unique risks that arise at the intersection of the virtual and physical worlds. Acknowledging this context, we aim to provide a solid foundation upon which future work can build. 

In the coming years, policymakers may quickly become aware of the risks posed by EAI because of headline-grabbing breakthroughs or threats from EAI. This could rapidly elevate EAI regulation on policy agendas, so policymakers must be equipped with appropriate tools and contextual understanding to create clear and beneficial legislation. Already, the overlap between the EU's AI Act and Machinery Regulations, which target robots, creates confusing and tangled requirements. Further policy action without regard for existing frameworks could impede, rather than improve, EAI safety. To ensure the safe and beneficial development of this transformative technology, we argue that policymakers must urgently build upon and address gaps in existing frameworks for robotics, autonomous vehicles, and agentic AI.

\section{Taxonomy of Risks from EAI}

Drawing on existing research and current predictions about EAI trajectories, we identify and explore four crucial areas of EAI risks: physical, informational, economic, and social (see Figure 2). This taxonomy leads to our discussion of how existing policy frameworks  address—or fail to address—these EAI risks.

\subsection{Physical risks}

\paragraph{Purposeful or malicious harm} EAI systems present distinct physical risks due to their embodiment in the physical world. EAI technologies have already been designed and deployed with lethal intent, such as AI-controlled drones~\cite{NYTimes2023AIDrones,Bondar2025UkraineAIWarfare}.  However, fully autonomous military robots, often integrated with bespoke AI architectures~\cite{Demarest2025ScoutAI,Hambling2024RobotDogs}, are not yet widely used in combat. While highly or fully autonomous warfare is distinctly possible in the future~\cite{cummings_artificial_2017}, immediate risks arise from commercially available EAI systems, including AI-controlled quadrupeds and autonomous driving assistants.  Recent research has demonstrated that these systems inherit \textit{jailbreaking} vulnerabilities from LLM-based AI models~\cite{zou2023universal,chao2023jailbreaking,chao2024jailbreakbench,robey2023smoothllm}. This could allow malicious actors to subvert safety guardrails and perform a range of harmful and irreversible physical tasks, including detonating explosives and deliberately causing human collisions~\cite{robey2024jailbreaking,zhang2024badrobot,ravichandran2025safety}.  VLAs exacerbate this risk: an attacker might craft a visual scene or textual instruction that, when interpreted through a language-action policy, yields physically dangerous instructions not anticipated by vision- or language-only defenses~\cite{jones2025adversarial,wang2024exploring}.

\begin{figure}
    \centering
    \includegraphics[width=1\linewidth]{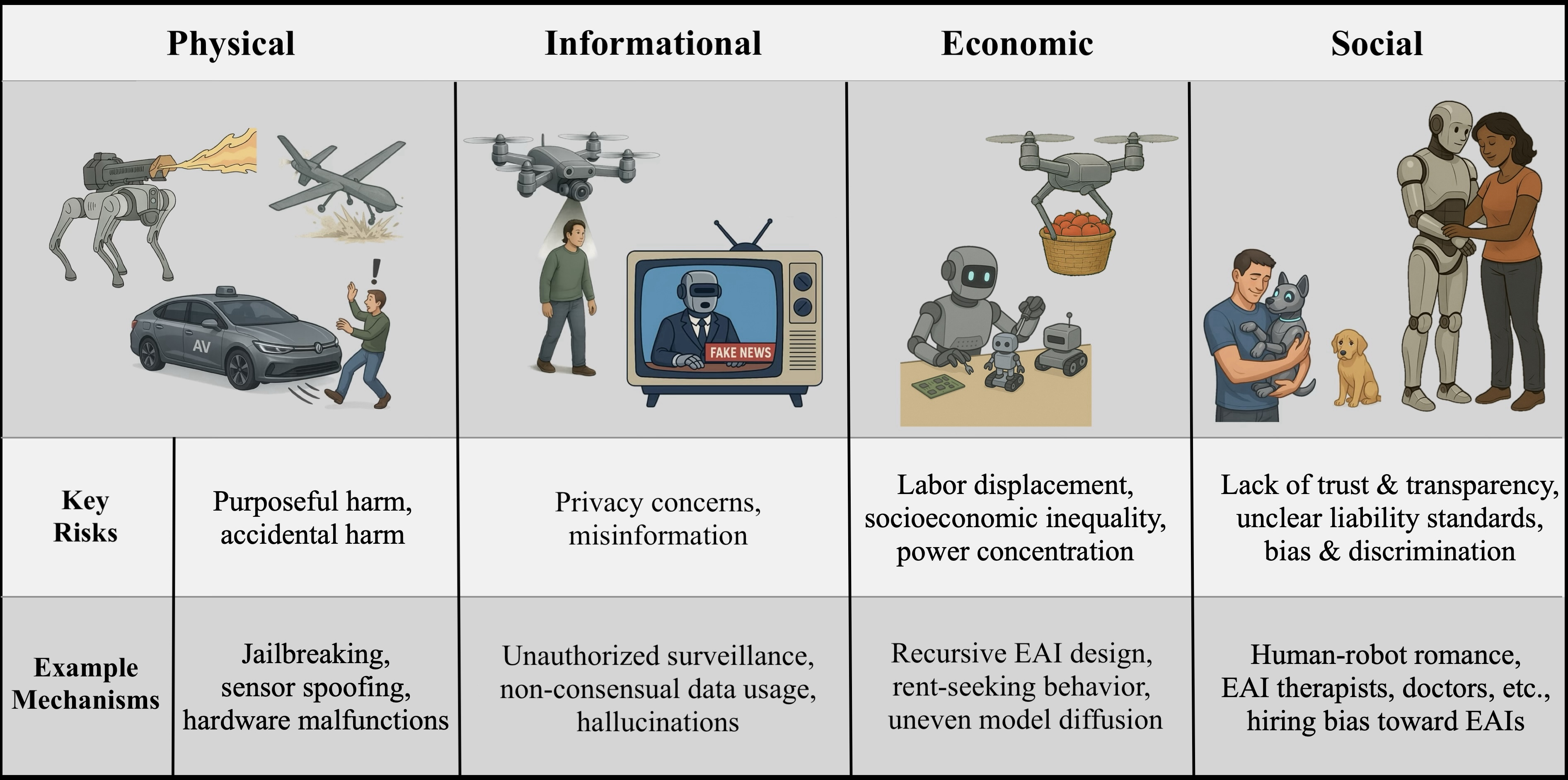}
    \caption{\textbf{An illustrated summary of risks from embodied AI}. We identify four key risk categories and provide several existing or potential mechanisms through which EAI systems could cause harm within each risk area.}
    \label{fig:enter-label}
\end{figure}

\paragraph{Accidental harm} Automation in sectors ranging from manufacturing to healthcare has and will increasingly put humans into close contact with EAI systems~\cite{garlick2024airobots}. This interaction increases the risk of accidental physical harm.  Though accidental harm has been a longstanding issue in industrial robotics, increased AI capabilities could exacerbate this risk; several recent reports document an increase in industrial injuries following the introduction of AI-controlled robots~\cite{atkinson2023robotdeath,fortune2023teslarobotattack,youn2018amazonbearspray}.  Virtual AI applications can also cause harm (e.g. through misinterpreted goals or misaligned behavior) but not directly in the physical world, unlike EAI. EAI's potential to cause accidental, physical harm could be caused by misspecified goals, lack of semantic understanding, misaligned behavior, physical hardware malfunctions, or other unanticipated behaviors~\cite{amodei_concrete_2016, ji_ai_2025,matheson2019human}. For example, a humanoid EAI might not correctly reason that placing a full glass of milk on a tilted table is perilous and likely to lead to a dangerous broken glass~\cite{fry_redefining_nodate}, or a swarm of EAI systems might get caught in a physical logjam and run over humans by mistake while trying to unstick themselves~\cite{winfield_ethical_2025}. Researchers also face persistent difficulty in getting models trained in purely virtual simulations to act as intended in the real physical world—what is referred to as the ``reality gap.''~\cite{Bousmalis}. This introduces significant scope for accidental physical harm if the deployed world does not closely match an EAI's training data~\cite{kaushik_safeapt_2022}.      

\subsection{Informational risks}

\paragraph{Privacy violations} EAI systems interact with huge amounts of data, creating significant privacy concerns. These systems are often trained on vast corpora and process a variety of data modalities---spanning visual, auditory, and tactile information---during deployment~\cite{kim_openvla_2024}. Like text-based virtual AI models, which are known to memorize and expose personally identifiable information~\cite{nasr2023scalable,carlini2019secret}, commercial robots have been shown to disclose proprietary information through simple prompts~\cite{robey2024jailbreaking}. Whereas virtual AI systems are constrained to collect data from either virtual interfaces or fixed points in the physical world (e.g. security cameras collecting facial-recognition data), EAI's mobility and the vast array of sensors used in EAI technologies expand concerns about unauthorized data collection. For example, EAI systems can monitor user behavior, infer physical preferences, and potentially contribute to future model training without the informed consent of those being observed beyond the limitations of immobile microphones or security cameras~\cite{calo_boundaries_2011,chatzimichali2020toward,guo_roomba_2022}. Bad actors within governments or corporations could gain access to private data streams and monitor users' movements 24/7, providing significant leverage over individuals to squash dissent or achieve personal power~\cite{davidson_ai-enabled_2025}. 
\paragraph{Misinformation} Non-embodied AIs are known to propagate misinformation~\cite{chen_can_2024,pan2023risk}.  Various studies have shown that LLMs hallucinate information, including academic citations~\cite{agrawal2023language}, clinical knowledge~\cite{singhal2023large}, and cultural references~\cite{weng2024hallucination}. EAI systems inherit these shortcomings in the physical world, answering user questions with deceptive or incorrect information~\cite{danaher_robot_2020}. Because VLAs fuse vision and language, their hallucinatory failures can be spatially grounded—e.g., misidentifying an object in view and then generating a plausible yet unsafe action plan around it. And although automated home assistants like Amazon's Alexa already lie about issues as innocuous as Santa Claus' existence~\cite{withers_alexa_2018}, more mobile, capable, and trusted EAI systems in sensitive positions (like home-assistant or community-service positions) could easily spread model developers' propaganda and talking points to users. For example, an EAI running on DeepSeek's latest model could provide a subtle yet continuous stream of misinformation to American users while performing tasks as innocuous as folding laundry or helping to cook dinner~\cite{myers_deepseeks_2025, lin_robots_2024}. 

\subsection{Economic risks}

\paragraph{Labor displacement} While virtual AI applications will likely displace certain types of human cognitive labor, EAI systems could significantly replace or displace physical human labor~\cite{du_humanoid_2024}. At a minimum, EAI will likely augment the type of work that humans perform~\cite{ACEMOGLU20111043, loaiza_epoch_2024}. Classical industrial robots have taken over many human roles in manufacturing~\cite{Industry}, and research has shown that robot deployment can lead to a reduction in human employment~\cite{acemoglu_robots_2020}. Future technological advances will likely accelerate this displacement process, as increasingly capable EAI systems perform complex, multi-step physical tasks beyond assembly lines—for instance, by serving as tourist guides or teaching in classrooms, and all without the need for sleep, breaks, sick leave, or vacation~\cite{newton_humanoid_2019}. Though automation has historically redirected labor toward areas of human comparative advantage~\cite{acemoglu_automation_2019}, AGI-enabled EAI could potentially automate all physical labor~\cite{bullock_preparing_2023}.

\paragraph{Socioeconomic inequality} Along with displacing labor, EAI could significantly exacerbate wealth inequalities. Those who have access to or own EAI systems will be able to automate labor and perform many tasks significantly better or faster than those without access. These significant productivity advantages will potentially concentrate wealth and exacerbate domestic and international inequality~\cite{costinot_robots_2018, korinek_artificial_2017}. For example, while a wealthy businesswoman could invest in a fleet of the latest humanoid robots, individuals lacking adequate capital might be forced to rent their EAI systems. This division could create stark and entrenched socioeconomic divides as the importance of outsourcing labor to ever-more-capable EAI increases~\cite{freeman_who_2016}. Virtual AI applications may cause similar socioeconomic inequality, but the ability to use and control access to EAI systems may confer significant and unique returns on investment, given that many tasks in the physical world necessary for human survival (e.g. growing food, building shelter) are constrained by human strength and energy.   

\paragraph{Power concentration} EAI deployment could accelerate the consolidation of economic and political power. Unlocking increasing returns to capital for EAI owners, EAI will decrease employers' reliance on and responsiveness to the needs of human labor~\cite{kulveit_gradual_2025}. Further, EAI users or consumers may become dependent on EAI owners for goods and services due to the productivity advantages conferred by EAI systems~\cite{berg_should_2018, ford_rise_2016}. The importance of EAI to perform physical tasks will likely exacerbate power-concentration risks presented by purely virtual AI systems. The proliferation of EAI systems could thus lead to a rapid concentration of corporate economic (and social) power, potentially even facilitating an eventual coup involving EAI~\cite{widder_why_2024, davidson_ai-enabled_2025}.  

\subsection{Social risks}

\paragraph{Bias and discrimination} Like virtual applications of AI, EAI can display bias towards and discriminate against users. When EAI systems are placed in positions of power, their biases could have significant impacts on fairness in everyday interactions and on general social dynamics~\cite{howard_ugly_2018, londono_fairness_2024}. For example, a peacekeeping humanoid robot may discriminate based on skin color~\cite{azeem_llm-driven_2024}. Unlike virtual AI applications, this bias can have immediate and irreversible physical consequences (e.g. if the peacekeeping robot mistakenly injures an innocent passerby). 

\paragraph{Lack of accountability and liability} Determining responsibility when EAI causes harm requires new accountability and liability frameworks that address the complexities of highly autonomous physical systems. Human users may disagree with decisions taken by expert EAI systems, raising significant questions of delegation and responsibility~\cite{calo_delegation_2016}. Lack of EAI accountability could lead to confusion for users and breakdowns in traditional justice systems~\cite{rachum-twaig_whose_2020}. For example, we may soon need to consider who to blame and how to collect damages when a highly autonomous robotic surgeon removes a healthy organ by mistake~\cite{Guerra_Parisi_Pi_Seidel_2024}. Although virtual AI applications also raise liability concerns, EAI's ability to cause physical damage underscores the importance of establishing robust liability regimes, as liability is crucial in remedying physical harms. 

\paragraph{Lack of transparency, explainability, and trust} Understanding how AI reaches conclusions or why AI systems perform specific actions motivates an entire  branch of interpretability research~\cite{olah_building_2018}, but physical embodiment raises the stakes for understanding these systems. For example, transparency of planned actions and explainability of decision-making is crucial when an AV suddenly changes lanes. A lack of transparency and explainability could lead to a lack of trust, which could become a critical and socially destabilizing issue with the widespread deployment of EAI~\cite{kok_trust_2020, hancock_meta-analysis_2011,ESTERWOOD2023107658}. 

\paragraph{Unhealthy or dangerous human-EAI relationships} Constant access to and interaction with EAI systems could foster dangerous human dependence or romantic attachment~\cite{rabb_attachment_2022}. People may depend on EAI systems for physical pleasure~\cite{cox-george_i_2018}. The physical presence and human-like features of EAI systems may significantly amplify the dependency issues already observed with conversational AI~\cite{fang_how_2025, freedman_day_2025}. People may easily fall in love with EAI systems, only to be distraught when these systems are altered or have their memories reset~\cite{jonze_her_2013}.

\paragraph{Transformative effects} EAI deployment could fundamentally reshape society, particularly if the speed of technological development outpaces society’s ability to adapt~\cite{ford_rise_2016, brynjolfsson_second_2014}. For example, EAI systems could provide physical threats of violence and mass surveillance capabilities to back up AI-enabled authoritarianism~\cite{barez_toward_2025}. Businesses might only employ EAI systems, leaving humans free to engage in other activities but also affecting how humans find meaning in their work~\cite{nikolova_robots_2024}. Humans also might lose the ability to perform various tasks as responsibilities are increasingly outsourced or delegated to EAI systems~\cite{kulveit_gradual_2025}—as an existing example, humans can lose natural abilities to navigate their environment when they outsource navigation to automated GPS systems~\cite{dahmani_habitual_2020}. Of course, purely virtual AI may also transform human society (and, to some extent, already has). However, it is difficult to comprehend how profoundly and completely EAI could alter how humans work, socialize, and structure our societies—for example by revolutionizing physical labor norms, automating self-improvement and production, or becoming humans’ primary social and physical companions~\cite{boyd_technology_2018}.

\section{Heat map of relevant policies}

Existing policy frameworks already address many risks identified in Section 2. Understanding how current regulations apply—or fail to apply—to EAI systems is essential for both policymakers and researchers. This section examines key legislation from the United States~(US), the United Kingdom~(UK), and the European Union~(EU) that governs related technologies, including classical robotics, autonomous vehicles, and virtual agentic AI. Our analysis identifies regulatory gaps specific to EAI by examining where existing frameworks provide minimal, adequate, or substantial policy coverage. This review, while not exhaustive, focuses on civilian applications of EAI.

\begin{table*}[t!]
\centering
\caption{\textbf{A summary of coverage of policies for major EAI risks.} 
We examine whether existing policies or governance frameworks exist to address risks from technologies related to EAI. 
\bcircle{} indicates that there is already a high level of coverage of relevant policies; \wcircle{} indicates there is partial coverage but that significant adjustments are likely necessary; \ecircle{} indicates a significant lack of governance frameworks to address the relevant risk. 
We reference AVs in particular rather than broader EAI, as most EAI regulations to date have addressed AVs.}

\label{tab:policies}
\begin{adjustbox}{width=1\textwidth}
\begin{tabular}{p{2cm}p{4.5cm}p{2.1cm}p{2.1cm}p{2.1cm}}
\toprule
\textbf{Risk} & \textbf{Subrisk} & \text{Classic robots} & \text{AVs} & \text{Virtual agents} \\
\midrule
Physical & Purposeful or malicious harm & \bcircle & \wcircle & \wcircle \\
& Accidental harm & \bcircle & \bcircle & \wcircle \\
Informational & Privacy violations & \wcircle & \wcircle & \wcircle \\
& Misinformation & \wcircle & \ecircle & \wcircle \\
Economic & Labor displacement & \ecircle & \ecircle & \ecircle \\
& Socioeconomic inequality & \ecircle & \ecircle & \ecircle \\
& Power concentration & \ecircle & \ecircle & \ecircle \\
Social & Bias and discrimination & \wcircle & \wcircle & \wcircle \\
& Accountability and liability & \wcircle & \wcircle & \wcircle \\
& Trust and transparency & \wcircle & \wcircle & \wcircle \\
& Human-EAI attachment & \ecircle & \ecircle & \ecircle \\
& Transformative societal effects & \ecircle & \ecircle & \ecircle \\
\bottomrule
\end{tabular}
\end{adjustbox}
\end{table*}

\subsection{Key policies}
The most robust existing policies and frameworks relevant to EAI concern physical and informational risks. This section first addresses several policy frameworks that govern physical harms, focusing on laws and standards that apply to AVs and robots given the lack of existing rulemaking for virtual agents. It then examines the major pieces of legislation that address informational harms. This section provides a brief overview of policies that address economic and social EAI harms given the limited existing policies in this area.

\paragraph{Physical risks} Emerging approaches to governing physical risks from EAI primarily target AVs and drones. AV-specific laws usually follow one of two approaches: adapting conventional automobile laws or creating bespoke legislation~\cite{geistfeld_roadmap_2017, kubica_autonomous_2022}. For example, the UK's Automated Vehicles Act 2024 introduced the concept of the Authorized Self-Driving Entity (ASDE) and the No-User-in-Charge (NUiC) operator. These entities (typically manufacturers or fleet operators) assume legal liability when the vehicle is in self-driving mode, effectively standing in for the “driver”~\cite{noauthor_automated_2024}. These ASDE and NUiC entities serve as helpful precedents for other forms of highly autonomous EAI; however, these roles were likely easier to introduce given continuity from other automotive regulatory efforts. Different forms of EAI—such as home care or educational EAI—will not have the same pre-existing foundation.  

Instead of creating AV-specific legislation, many EU countries largely govern AVs using existing product liability paradigms that hold manufacturers accountable when systems cause harm while operating within intended parameters (i.e. the operational design domain). However, these frameworks face challenges when applied to autonomous systems. For example, German liability law states that humans are still considered drivers when operating an AV with SAE International Level 3 autonomy, in which humans must take over the operation of the vehicle when the system requests it but otherwise do not manipulate the vehicle. In SAE Level 4 vehicles, which require minimal human input, an owner-like entity, referred to as a “keeper” of a vehicle is instead held strictly liable for injuries~\cite{steege_liability_2024}. This reliance on existing product liability law could help govern some forms of EAI, but directly charging the ``keeper'' of a model in the event of loss of life seems to overlook the complex web of relationships between the original manufacturer, software updater, operator, and owner. Stringent reliance on product liability law could also deter beneficial EAI development and deployment; manufacturers might be reluctant to risk exposure to judicial uncertainty as courts attempt to reconcile existing laws with increasingly intelligent and autonomous systems.    

In the United States, the recently proposed ADS-equipped Vehicle Safety, Transparency, and Evaluation Program (AV STEP) would require approved AV manufacturers to share details about their vehicles’ development and operation. This would include information about the types of simulations used to train the vehicles’ algorithms, what sorts of environments the vehicle is meant to operate within, and relevant vehicle oversight mechanisms to ensure operational safety~\cite{noauthor_ads-equipped_2024}. AV STEP is a promising framework that could be extended to other EAI contexts, though it remains unclear which regulatory body would oversee other modalities of EAI. The National Institute of Standards and Technology's Risk Management Framework could provide a helpful starting point for a sweeping, interagency approach to physical safety certification~\cite{joint_task_force_transformation_initiative_risk_2018}.

Laws concerning aerial drones are also relevant to EAI policy, although many of today’s drones do not involve AI and operate with limited autonomy~\cite{nawaz_regulating_2024}. For example, EU regulation on drones distinguishes between remotely-piloted drones, autonomous drones—which can navigate dynamically without pilot intervention—and automatic drones that instead fly pre-planned routes~\cite{EU_Drone}. The EU requires that all drones are operated by a remote pilot who assumes responsibility for each flight. Pilots of autonomous and other high-risk drones must pass theoretical-knowledge and practical-skill courses. However, National Aviation Authorities often rely on self-declarations from pilots that their operations pose minimal threats to nearby people, for example by not flying over groups of people ``uninvolved'' in the drone's operation. Non-autonomous drones require less stringent training to operate but face substantial operating restrictions; for example, pilots of these drones must maintain the drone in their line of sight at all times.

The UK’s regulation largely mirrors the EU’s drone rules, including a mandate that a drone be able to safely return to the ground in the event of a system failure or loss of communication~\cite{noauthor_uas_2025}. The drone regulation environment in the US is quickly evolving due to a June 2025 Executive Order that, among other changes, will make it easier for drones to operate beyond pilots’ line of sight for commercial and public-safety operations~\cite{unleashing_2025}. The US federal government requires that all drones are equipped with remote-identification capabilities and must be able to broadcast the drone's identity, location, altitude, and more~\cite{Part_2021}. The UK will implement similar tracking requirements in 2026~\cite{CAA_2025}.      

Another key piece of legislation is the EU's Machinery Regulation (MR), which was passed in 2023~\cite{noauthor_regulation_2023}. Updating similar legislation from 2006, the MR regulates many types of robots in the EU and explicitly addresses aspects of AI and EAI safety. Tobias Mahler thoroughly investigates how the MR interacts and overlaps with the EU's AI Act~\cite{mahler_smart_2024}, identifying the MR's attempt to future-proof its regulations through mentions of machines with ``self-evolving behaviour...designed to operate with varying levels of autonomy.'' The MR encompasses a comprehensive spectrum of physical safety concerns, ranging from the materials used to emergency-stopping systems to the risk of being trapped inside a machine. The MR mandates that machines sold in the EU must be tested for compliance with these safety regulations; as with other EU regulations, third-party evaluators (or ``notified bodies'') test whether machines fulfill the safety requirements. 

Beyond legislation, international standards provide manufacturers and deployers with robust safety guidance for robotics, primarily focused on physical safety. For example, ISO 10218:2025 recommends safety protocols for assessing risk, mitigating risk (e.g. through controls, safety functions, stopping functions), and safe-design certification for industrial robots~\cite{noauthor_robotics_2025}. In addition, ISO 13482:2025 addresses safety requirements for service robots in personal and professional settings, emphasizing sensor reliability, uncertainty management protocols, and decision verification through multiple sensing modalities~\cite{noauthor_robotics_2025_safety}. Many standards also exist explicitly for AVs. For example, ISO/SAE 21434:2021 and UN Regulation 155 provide standards for cybersecurity engineering for road vehicles~\cite{noauthor_road_2021, noauthor_cyber_2021}. These cybersecurity standards are particularly relevant for many forms of mobile EAI, especially given that EAI systems may be deployed in situations without regular access to trusted networks~\cite{thakker_risk-guided_2025} (e.g. during disaster rescue missions, or when operating with law enforcement in hostile territory).   

\paragraph{Informational risks} Many informational risks apply to both virtual and embodied AI applications. Key frameworks governing EAI informational risks include the EU's AI Act and the General Data Protection Regulation (GDPR). This legislation prohibits several data practices relevant to EAI, such as untargeted facial image scraping and manipulative decision-influencing techniques~\cite{noauthor_artificial_2024}. These restrictions are crucial for EAI systems that continuously collect data in public and private settings~\cite{Mintrom}.

GDPR legislation in the EU and UK establishes strict governance requirements concerning the capture, use, and storage of data. GDPR requires that data is only collected for ``legitimate interests'' and that entities collecting data are classified as ``data controllers'' who must document data collection practices, usage patterns, and storage methods while implementing robust security measures~\cite{GDPR}. These stipulations address many concerns with the privacy and treatment of data collected by EAI systems. However, EAI deployments in public spaces challenge traditional consent models and controller identification~\cite{Mintrom}. Notions of data controllers and implied consent to be recorded will need to be substantially altered or clarified with EAI deployment. EAI systems deployed in public spaces, for example, raise questions about how to opt out of data collection, understand who receives and controls data collected from the system's sensors, and even what constitutes a public vs. private space~\cite{barfield_robots_2024}. 

\paragraph{Economic risks} Regulatory frameworks addressing the economic impact of EAI remain underdeveloped. Existing legislation, such as the UK's Employment Rights Act 1996 and the US WARN Act, provides limited protections for workers facing technological displacement~\cite{noauthor_employment_1996, noauthor_worker_1988}. Labor organizations have achieved isolated victories against automation, exemplified by the International Longshoreman Association's recent successful challenge to port automation~\cite{berger_port_2025}. Policymakers must better prepare for widespread labor displacement resulting from the deployment of EAI in industrial and service settings. In reordering and reimagining labor structures, EAI innovation could lead to a period of rapid creative economic destruction~\cite{schumpeter_capitalism_2008}.

Some observers think AI development represents a different kind of technological transition compared to previous transformations, as AI may replace cognitive tasks in addition to physical labor~\cite{suleyman_coming_2023, cheng_forging_2025}. Economic policies may not need to target technological failures, as with many physical and informational risks mentioned in Section 2. Economic risks may instead emerge because EAI systems work ~\textit{too} well and rapidly upend the need for human labor. As a result, policymakers should consider social policies to manage these emerging tensions~\cite{mokander_artificial_2024}.

\paragraph{Social risks} Few regulations directly address the social impacts of EAI. Those that do exist largely govern issues of direct human interaction with EAI systems and do not address larger issues of how society will transform as these entities become increasingly prevalent and powerful. The EU AI Act's broad prohibition on infringing fundamental rights could be extended to address issues surrounding a lack of trust, lack of transparency, unhealthy or dangerous attachments, and bias and discrimination, but this would require further specification. In terms of accountability, proposed frameworks for attributing actions and delegating authority to virtual agents could prove helpful for EAI~\cite{chan_infrastructure_2025, south_authenticated_2025}. 

GDPR Article 22 provides an instructive example of existing regulation that implicates but does not directly address EAI systems. Likely designed with virtual AI applications in mind, the Article prohibits individuals from being ``subject to a decision based solely on automated processing'' when that decision has legal or similarly significant consequences for that individual~\cite{GDPR}. However, it remains unclear how this Article could be reconciled with fully autonomous EAI, or how individuals could appeal to a human intervener—as the Article later mandates—in immediate physical interactions or conflicts with EAI systems. 

Beyond legislation, several international standards aimed at manufacturers and developers emphasize transparency, ethical design, and trustworthiness in EAI systems. However, these standards, including the IEEE’s 7000 series on autonomous system transparency~\cite{IEEE}, algorithmic bias~\cite{IEEE2}, and the impact of robotics on human well-being~\cite{IEEE3}, are voluntary. These standards could apply to a wide variety of EAI instantiations or applications, but their voluntary nature would similarly limit their impact. 

\subsection{What are the most significant gaps?}

Though major building blocks to address harm from EAI systems already exist, several key policy gaps concerning EAI safety require urgent attention. 


First, there are significant gaps concerning robust certification processes for different EAI modalities. Regulating AVs is straightforward in many senses due to their defined operational domains (e.g. cars usually stay on roads). The EU's MR mandates certification of safety for a broader array of EAI modalities (with carve-outs for AVs or aerial drones) and stipulates that EAI ``shall not...perform actions beyond its defined task and movement space.'' Current drone regulations address many aspects of aerial drones, but this regulation largely focuses on drones lacking true autonomous navigation capabilities. Future instantiations of EAI, however, will likely have significantly expanded freedom of movement, enabling them to enter private residences and commercial establishments and conduct surveillance in schools or public areas. These expanded domains require thoughtful processes to certify the operational safety of EAI systems—such frameworks do not yet exist. Expecting existing consumer-safety laboratories, which currently test the safety of machine components like materials and locking mechanisms, to evaluate the safety of EAI sytems is unrealistic. Basic questions such as identifying the relevant regulator are a key starting point—to what extent should there be EAI-specific consumer protection boards with AI expertise, or should existing third-party testing laboratories take on this responsibility? 

Secondly, once a suitable apparatus is in place, EAI capabilities should be measured with reliable and valid evaluations and benchmarks. To date, few of these benchmarks exist~\cite{sermanet_generating_2025}, despite the existence of a range of benchmarks for virtual AI systems~\cite{noauthor_preliminary_2024}. Voluntary standards can provide technical guidance, but the lack of laws enforcing benchmarking and evaluations across all four key risk areas (physical, informational, economic, and social) is a critical policy gap. Evaluations could cover a range of considerations outlined in our taxonomy, for example evaluating the robustness of simulation-to-real protocols (the ``reality gap''), the conformity of EAI systems with their stated operational domain, robustness to jailbreaking, cybersecurity measures, alignment between software and hardware capabilities, and hardware durability and reliability, among other areas.

Thirdly, policies or frameworks currently devoted to post-deployment EAI monitoring are unclear or lacking in detail. These sorts of oversight and monitoring mechanisms have been highlighted for other AI systems~\cite{dunlop_safe_2024}. However, current regulations requiring EAI systems to include ``black boxes'' that record and preserve data in the event of accidents, crashes, or misuse are hazy. The EU's MR mandates that data about safety-related decision-making processes is kept for a year after its collection~\cite{noauthor_regulation_2023}. At the same time, the EU's AI Act provides contradictory guidance that high-risk AI systems must retain this information for at least six months~\cite{noauthor_artificial_2024}. These types of recording systems, in addition to live data monitoring, can enhance system safety and aid post-incident investigations~\cite{9721088, mathew_who_2025}. The EU's MR also states that ``it shall be possible at all times to correct the machinery...to maintain its inherent safety,'' but this notion of oversight requires significant clarification for highly- or fully-autonomous systems. Does this involve the ability to tweak EAI actions in real-time, send new model updates over the air, or some other intervention? Who is conducting this monitoring—users, the government, or a private, delegated oversight entity (perhaps AI-driven itself)?

There are likewise stark gaps in policies addressing economic and social risks from EAI. Although EAI could cause mass labor displacement, proposals to distribute economic benefits are still in their infancy~\cite{yelizarova_missing_2025, merola_inclusive_2022}. EAI could lead to many positive economic and social outcomes as well, but policymakers must then ensure that economies around the world have the sovereign systems (data centers, energy production, EAI hardware) necessary to seize EAI's benefits. No well-articulated policy on this, whether national or regional, yet exists. Similarly, policies addressing social issues related to trust and human-EAI attachment are currently scant~\cite{bao_mitigating_2023}. More broadly, there are significant policy gaps at the intersection of EAI and AGI. For example, should an EAI system be allowed to build other EAI systems? Should a country developing AGI in embodied form automatically and freely share the technology with the rest of the world? There has likewise been little policy attention devoted to EAI defensive acceleration. If AGI is as powerful as some observers imagine, it seems possible or even plausible that AGI could help solve a raft of open governance questions and issues, particularly in the defense arena~\cite{clifford_introducing_2024}. Yet what does it mean for a society to be protected by an army of EAI systems? Policymakers must urgently consider how EAI should be developed, deployed, and integrated into societal structures to address the broad array of currently neglected challenges mentioned here.

\section{Proposed pathways forward}

To effectively mitigate these urgent EAI risks, ensure beneficial EAI development, and create a balanced regulatory environment, policymakers must fill gaps in today's fragmented policy landscape with pragmatic approaches that adapt to the complexity of emerging EAI technology.  

\subsection{Invest in EAI safety research}

Based on the risk taxonomy described in Section 2, we recommend significant and increased research be devoted to EAI safety. For example, robotics and machine-learning researchers can further efforts to make hardware actuators less susceptible to hacking and malfunction through physical design and formal methods~\cite{basin_formal_2021}. Building benchmarks and evaluations of EAI capabilities and behavior is a particularly promising area of EAI safety research. Most AI benchmarks and evaluations today specifically target the virtual aspects of AI~\cite{grey_safety_2025}, although recent progress has been made in EAI-specific research, such as at Google DeepMind~\cite{sermanet_generating_2025}. Researchers should and build on this progress by developing EAI evaluations and benchmarks that span a broad set of tasks and task types~\cite{yin_safeagentbench_2025, choi_lota-bench_2024}, similar to the work being done by the RoboArena team~\cite{atreya_roboarena_2025}. For example, specific attention could focus on ensuring the safety of EAI systems acting in multi-agent systems or swarms. EAI systems will almost certainly navigate complex environments with others that do not share the same goals, so inter-agent collaboration and coordination are paramount to avoid poor outcomes, as recently highlighted by the World Economic Forum and researchers from Cooperative AI~\cite{larsen_navigating_2024, hammond_multi-agent_2025}. Beyond physical risks, benchmarks and evaluations should also address issues related to privacy and cybersecurity, for example by building upon zero-knowledge proof research from other AI domains~\cite{peng_survey_2025}. We also need benchmarks that stress-test the joint vision–language–action loop—measuring, for instance, whether a VLA model’s visual prompt leads to safe, context-aware behavior across edge cases. Benchmarks and assessments will not address every risk raised above—particularly socioeconomic considerations—but they are a critical step towards minimizing many risks from EAI. 

Structuring and operationalizing increased EAI safety research demands particular attention and governance efforts. Some research efforts could likely be integrated into existing national AI Safety Institute initiatives, however, other safety research will likely be sector-oriented (e.g. healthcare, construction, education, etc.) and may be best directed through existing industry regulators. Deciding which research efforts to prioritize, how to disburse funding, and ensuring that dispersed research efforts complement each other all require further consideration.  

\subsection{Create robust certification requirements before EAI deployment}

National bodies should mandate that EAI systems pass safety evaluations and are certified for public use. EAI systems should have clear ‘model cards’ describing how they were trained (e.g. what sorts of data were used, how a model performs on safety benchmarks), in which domains it was designed to operate, and what safety measures the manufacturer has taken to ensure its safe operation. If desired, policymakers could then mandate that EAI systems be limited to legally operating within the specified domains (as in the EU's MR), potentially aided by remote identification requirements similar to those currently applicable to drones. This model card approach would borrow from the frontier safety frameworks that many leading AI labs have implemented~\cite{buhl_emerging_2025}. This regime could be enforced via audits of EAI manufacturers and developers~\cite{mokander_auditing_2023}. Policymakers should clarify which entity is responsible for verifying and validating EAI safety, possibly by establishing a national laboratory for EAI testing as part of existing AI Safety Institutes or by assigning this responsibility to private-sector actors. For a stricter approach, policymakers could also decide that only EAI systems quantifiably proven safe, as currently being investigated in conjunction with the UK's ARIA funding agency, can be deployed~\cite{dalrymple_safeguarded_2024}. Policymakers should also ensure that this certification regime incorporates different categories of requirements based on potential risk, as risk from EAI depends heavily on the sensitivity of an EAI's deployed context and its capabilities, as noted in the recent RAND report ``Averting a Robot Catastrophe.''~\cite{vermeer_averting_2025} For example, certifying EAI safety for an EAI version of a children's toy and a autonomous limousine should involve different safety testing requirements and thresholds. Designing the exact certification regimen and constructing these different categories will require significant collaboration with technical experts. Policymakers should ensure that this sort of regulation is reasonably limited in scope and supports beneficial innovation while mitigating risks.

More broadly, EAI regulation should address concerns at the developer, model, and application layers, much like approaches for non-embodied AI~\cite{mokander_auditing_2023}. Each of these layers best identifies and manages different policy issues and ethical and social risks. This proposed certification scheme for EAI could address concerns at the model and application levels unique to EAI, like considerations about the simulation-to-real gap and hardware-software compatibility issues. These concerns likely would not be covered by existing policy efforts focused on non-embodied AI. Combining this model- and application-specific approach with policy efforts at the developer layer could help ensure robust and durable EAI safety.

\subsection{Promote industry-led standards to address EAI risks}

Industrial and standards bodies can push forward EAI safety efforts in tandem with legislative approaches. These standards bodies can develop robust updates to existing standards and create dynamic new standards that address the increased capabilities of EAI. Existing standards are grounded in today's robotic capabilities. Still, even recent updates fail to incorporate or address how highly or entirely autonomous robots capable of advanced reasoning will affect industrial and service applications. Notably, in May 2025 the ISO announced intentions to create a new standard for humanoid robots, which should encompass a range of additional form factors, autonomy, and use cases~\cite{noauthor_isoawi_2025}. These standards should address a range of technical protocols (e.g. cybersecurity and jailbreak-proof), while also mandating that EAI systems are equipped with tools to foster transparency and accountability. For instance, future standards should mandate that EAI systems are equipped with ``black boxes'' that record the system's sensor input and, if possible, its reasoning in the minutes preceding an adverse event. These black boxes will raise privacy concerns of their own, as they would create an extra attack surface for data exfiltration. The tension between accountability and privacy should be acknowledged and addressed in future discussions.

The fast-evolving nature of EAI also requires that standard-setting and evaluation regimes can adapt quickly to new technological developments. 
Industrial actors can leverage their technical expertise to help develop and adjust standards, which larger international standards-setting bodies might be too slow to enact~\cite{roberts_can_2025}. For example, deployment of EAI in dynamic, real-world situations makes it difficult to assess safety via classical approaches that involve repeatable and exhaustible safety evaluation. Instead, demonstrating EAI and robotic safety is increasingly reliant on statistical significance and repeatable showcasing of safe behavior~\cite{ASTM, weng_rethink_2025}. More broadly, these industrial standards should be developed across the entire EAI continuum—addressing aspects of the EAI stack from components to scenarios to systems, and swarms~\cite{liu_establishing_2024}.

\subsection{Clarify liability regimes for fully autonomous systems}

National and international policymakers should clarify existing, muddy liability regimes. When truly autonomous EAI systems are deployed, who should be held accountable for injuries or misuse? Should the person who gave the model its latest instruction be at fault, or should the blame rest with a software developer or the original manufacturer? When hardware and software diverge—e.g. a new LMM is released but runs on existing hardware—how does this change liability? Is the original manufacturer no longer at fault? In the future, how should EAI systems themselves be held accountable for faults if they are considered agentic? Unlike current AVs, many EAI systems will likely be fully automated (the equivalent of SAE Level 5). If true, full automation is reached, there will by definition not be humans in the loop to be held accountable or liable. EAI liability is a growing area of legal study~\cite{barfield_cambridge_2024, calo_application_2016, White2017, kolt_governing_2025}, but firm policies need to replace today's unclear and ad-hoc legal approaches. For instance, policymakers should clearly define notions like the Authorised Self-Driving Entity laid out in the UK Automated Vehicles Act to designate responsibility for EAI operation. Policymakers must recognize the tension made apparent in GDPR's Article 22 between recourse to human intervention and fully autonomous physical systems. For example, policymakers could prohibit EAI deployment in situations where recourse to human intervention or decision-making would not be logistically possible or instead update the GDPR to outline critical scenarios in which humans do not have recourse to this option. At the same time, policymakers must work with technologists to determine when a manufacturer should be held accountable for errant EAI actions (e.g. when training is deemed insufficient for deployment in specific environments, or when new models are released but manufacturers do not make safety-relevant over-the-air updates available). 

\subsection{Plan and prepare for the transformative economic and social effects of EAI}

Policymakers at the national and international levels should draft legislation to prepare safety-net or assistance programs for people whose labor is replaced by EAI systems. Basic proposals have been floated concerning UBI~\cite{okeefe_windfall_2020}, or even universal basic compute (UBC), whereby people are guaranteed access to and use of AI or EAI systems~\cite{varanasi_openais_2024}. However, these proposals remain very sparse and abstract. Policymakers should create draft frameworks and attempt to form early consensus now, as highly advanced EAI models and widespread labor displacement may arrive in the near future. Policymakers should specifically address who will be eligible to claim these social assistance packages and under what conditions (e.g. what type of proof will be required to demonstrate that an individual lost their job as a direct result of EAI automation). Reskilling programs are another potential policy avenue; however, these worker retraining programs may face limitations in the face of AI and EAI that automate an increasing number of jobs~\cite{jacobs_ai_2025}.

Similarly, policymakers must better prepare for transformative social effects~\cite{kulveit_gradual_2025}. Given the capital-intensive nature of EAI systems, it is plausible that EAI power and access could be concentrated in the hands of a select few~\cite{ford_rise_2016}. Policymakers should draft options to combat this social power concentration, perhaps through targeted taxation mechanisms~\cite{thuemmel_optimal_2023, ahn_navigating_2024, mazur_taxing_2019}. Policymakers should also fund research on how to mitigate adverse emotional dependencies between EAI systems and humans. EAI deployment is ultimately (for now) a human decision, so national and international policymakers should consider whether some domains should be entirely off-limits for EAI interaction. Organizations such as the OECD, GPAI, or the nascent UN AI Panel and Dialogue should prioritize action-oriented EAI dialogue on these pressing social issues, as EAI will impact people worldwide, not just in today's robotics hotspots.  
 
\section{Discussion and Limitations}

Our analysis has several limitations, for example regarding the geographies addressed and the type of EAI applications mentioned here. Furthermore, several key counterarguments to our main claims warrant further attention. We hope this discussion helps pave the way for future work on this exciting topic.

\paragraph{Geographical limitations} We focus on policies from the US, UK, and EU, given the authors' location and expertise. However, future analyses should extend to other geographies, particularly China, Japan, and India. China, for example, is leading many aspects of EAI production and is likely to debut some of the world's most comprehensive EAI regulations. For example, Chinese officials recently announced rules regulating over-the-air software updates for autonomous vehicles—one of our identified risk vectors—due to their potential to conceal defects or contain bugs~\cite{reuters_2025, reuters_2025_april}. It is also important to note that notions of safety depend in part on cultural and societal ideas that shift across geographies, so it is critical that conversations about EAI safety involve robust global representation and perspectives~\cite{okolo_re-envisioning_2025}.

\paragraph{Application limitations} This paper primarily focuses on civilian applications of EAI; however, military and law enforcement EAI applications also demand urgent policy action. As seen with the use of drones in Ukraine and Russia, EAI systems can act increasingly autonomously to inflict significant damage thousands of miles away from front lines~\cite{varenikova_ukraine_2025}. The development and deployment of weaponized EAI systems are expected to continue growing over the coming months and years. This could significantly lower barriers for non-state actors to cause significant damage and for political leaders to suppress dissent with fleets of embodied agents~\cite{davidson_ai-enabled_2025}.          

\paragraph{Methodological challenges} EAI is a vast, rapidly growing, and fast-evolving field. It would be nearly impossible to address every aspect of EAI progress or document every EAI safety concern in this piece. For pragmatism's sake, we omit many exciting areas of analysis and discussion. For example, the potential for high-fidelity virtual simulations and video-based learning to revolutionize EAI training merits its own review and policy recommendations~\cite{fry_redefining_nodate}. We also acknowledge that there is a vast and growing literature devoted to AI safety in general. Enhancing the safety of underlying LLMs and LMMs is key, and we address this point further below. We also recognize that the EAI field is inspired by decades of research from engineering, computer science, and human-robot interaction disciplines. Entire libraries are filled with information that is acutely relevant to EAI's contemporary development and informs policy reactions. With limited time and resources, we chose to focus our analysis on the less-explored crossover of contemporary, quickly advancing agentic AI and classical robots. 

We likewise had to make difficult decisions in creating our main categories of risk. These categories (physical, informational, economic, and social) are not absolute and sometimes overlap with each other (e.g. some economic risks, especially power concentration, can also be considered as a social risk). However, these categories continually appeared in our conversation with experts and in our literature review. We hope these categories serve as a starting point for more robust discussions about the key areas of EAI risk, which will in turn enable policymakers and practitioners to better mitigate emerging EAI threats.  

\paragraph{Market forces, incentives, or societal pressures} We recognize that good policy does not always mean more regulation; active intervention may not be necessary if market or social pressures naturally lead to safer EAI. For example, manufacturers or countries with the strongest safety protocols might be seen as having a commercial advantage, similar to how Apple has made privacy a key selling point~\cite{leswing_apple_nodate}. However, as with virtual AI systems, market forces may lead to race dynamics and exacerbate risks~\cite{GRUETZEMACHER2025103563}. EAI systems will have significant use cases, and the associated financial incentives to rapidly ship and sell products will likely translate to industry efforts to minimize government oversight at the expense of product safety. 

\paragraph{Technical solutions and EAI harm} Technical solutions (e.g., alignment, unlearning, etc.) will play a crucial role in mitigating risks associated with EAI and enhancing oversight and control. However, technical solutions to EAI risk may only be established and implemented effectively after EAI risks emerge at scale, as technical solutions are often responsive to demonstrated needs~\cite{mehrabi_survey_2022}. Technical solutions will also likely not address all the risks mentioned above—especially those related to economic and social risks. A more balanced and pragmatic approach is required, one that combines the best aspects of technical and non-technical solutions~\cite{watson_competing_2024}. 
    
\paragraph{Hardware and physical engineering limits} Hardware limitations constrain EAI behavior in ways that virtual agentic AI does not encounter. Finite battery capacity, limited on-device processing power, and curtailed range of motion are physical engineering problems unique to EAI~\cite{hecht_robots_2023, groshev_edge_2023, li_developments_2025}. These problems naturally mitigate risk to some extent, as EAI designers and creators must first attempt to overcome these physical limitations to achieve greater capabilities. This lends designers more control and time to create lower-risk designs in the short term, although these limitations may be circumvented or solved as research accelerates. If AI systems become involved in designing and testing new hardware systems, the length of feedback loops and the time required to ideate, manufacture, and implement more capable hardware will likely shrink.

\paragraph{Overlap of EAI risks with wider LLM risks} Skeptics could argue that many issues discussed in this paper are just inherent LLM risks. In other words, if we make LLMs safe, we will solve EAI risks. We acknowledge that there is considerable overlap between risks from AI and EAI~\cite{weidinger_taxonomy_2022}. However, EAI raises specific concerns about safety at the application and organizational level, in addition to the fundamental model level~\cite{mokander_auditing_2024}. Many risks are modified, magnified, or made more urgent by its presence in the physical world. EAI creates immediate risks (e.g. a malfunction during surgery could result in a patient's death), pervasive (e.g. with continuous presence in households, workplaces, and public areas), and growing with scale (e.g. swarms of public-safety EAI systems coordinating to enforce martial law). These risks will likely be exacerbated as EAI capabilities, autonomy, and deployment increase. For example, an EAI could impair a baby's healthy development~\cite{lin_rights_2012} or be hacked and cause a deadly crash in ways inapplicable to purely virtual AIs~\cite{yousseef_autonomous_2024}. Improving LLM safety alone is helpful but would fail to cover these—and many other—critical EAI scenarios. 
    
\section{Conclusion} 

The EAI field is rapidly advancing, driven by increasing hardware investment, breakthroughs in LLMs and LMMs, and quickening deployment. These trends will likely accelerate in the coming years. However, policymakers around the world have thus far neglected EAI governance. Frameworks governing EAI, where they exist, have hardly advanced even though associated risks have gradually transitioned from the realm of science fiction to the real world. 

When a sudden EAI disaster or ChatGPT-like breakthrough does happen, though, it would be misplaced and misguided for policymakers to reinvent the policy wheel. Policymakers must plug the gaps in existing frameworks where risks from EAI are insufficiently addressed, especially regarding policies for classical robots, automated vehicles, and virtual agentic AI. This matters today; we now have a crucial opportunity to create pragmatic EAI policies as EAI technologies are increasingly deployed and relied upon. 

We have argued that policymakers should encourage the development of effective benchmarks, evaluations, and safety protocols for the responsible deployment of EAI; ensure safety certification for a range of EAI form factors, capabilities, and operational domains; reevaluate liability paradigms; confront labor displacement; and address larger societal issues, such as human-EAI attachment. These recommendations aim to provide sensible first steps that can guide and encourage safe EAI innovation with minimal downside. Many other risks necessitate minor tweaks or adjustments to existing policies—for example, preventing privacy violations from EAI systems in public will require thoughtful integration with existing laws such as the MR and GDPR. Zooming out, research on EAI safety in general should be significantly expanded, and policymakers must collaborate with robotics and machine-learning researchers and practitioners to translate findings from academic and industry research into policy. EAI is rapidly advancing, and its risks are quickly becoming real; policymakers must urgently address policy gaps to mitigate these critical risks.  

\section{Acknowledgements} Jared Perlo is grateful for the support of the Centre for the Governance of AI and French Center for AI Safety (Centre pour la Sécurité de l'IA, or CeSIA) for making this research project possible. We would like to thank Aaron Prather, Aidan Homewood, Amelia Michael, Connor Aidan Stewart Hunter, Edward Kembery, Jenny Read, Keegan McBride, Krzysztof Bar, Kyler Zhou, Liam Patell, Markus Anderljung, Marta Ziosi, Nora Amman, Pierre Sermanet, Roeland P.-J. E. Decorte, Shaoshan Liu, Simon Mylius, Todor Davchev, Umair Siddique, Yohan Mathew, and Yunzhu Li for their valuable input and collaboration.

\newpage
\bibliography{bibliography}
\bibliographystyle{unsrtnat}

\end{document}